\documentclass[a4paper,10pt]{IEEEtran}

\pdfoutput=1
\IEEEoverridecommandlockouts
\makeatletter
\def\endthebibliography{%
	\def\@noitemerr{\@latex@warning{Empty `thebibliography' environment}}%
	\endlist
}
\makeatother

\IEEEoverridecommandlockouts
\usepackage{setspace}
\usepackage{multirow}
\usepackage{lipsum}
\usepackage{amsfonts}
\usepackage{array}
\usepackage{amsmath,cases,amssymb}
\usepackage{amsmath}
\usepackage{amsthm}
\usepackage{graphicx}
\usepackage{cite}
\usepackage{mathtools}
\usepackage{subfigure}
\usepackage{epsfig}
\usepackage{url}
\usepackage{algorithm}
\usepackage{algorithmic}
\usepackage{epstopdf}
\usepackage{color}
\usepackage{makecell}

\usepackage{lipsum}

\usepackage[justification=centering]{caption}

\DeclareGraphicsExtensions{.eps,.pdf}

\newcommand{\bPsi}{\boldsymbol{\Psi}}

\newcommand{\bw}{\mathbf{w}}

\newcommand{\bC}{\mathbf{C}}

\newcommand{\bH}{\boldsymbol{H}}

\newcommand{\bB}{\boldsymbol{B}}

\newcommand{\bI}{\mathbf{I}}

\newcommand{\bu}{\mathbf{u}}

\newcommand{\bP}{\mathbf{P}}

\newcommand{\bo}{\boldsymbol{o}}
\newcommand{\bb}{\mathbf{b}}
\newcommand{\ba}{\boldsymbol{a}}
\newcommand{\bs}{\boldsymbol{s}}
\newcommand{\bK}{\mathbf{K}}

\newcommand{\bmu}{\pmb{\mu}}

\newcommand{\Ptt}{\mathtt{P}}
\newcommand{\Ott}{\mathtt{O}}

\newcommand{\Scal}{\mathcal{S}}
\newcommand{\Acal}{\mathcal{A}}

\newcommand{\Qcal}{\mathcal{Q}}
\newcommand{\Kcal}{\mathcal{K}}

\newcommand{\Mcal}{\mathcal{M}}

\newcommand{\Xcal}{\mathcal{X}}
\newcommand{\Ncal}{\mathcal{N}}

\newcommand{\Pcal}{\mathcal{P}}
\newcommand{\Ocal}{\mathcal{O}}
\newcommand{\Tcal}{\mathcal{T}}

\newcommand{\PA}{\mathtt{PA}}
\newcommand{\SA}{\mathtt{SA}}

\makeatletter
\newcommand{\doublewidetilde}[1]{{%
		\mathpalette\double@widetilde{#1}}}
\newcommand{\double@widetilde}[2]{%
	\sbox\z@{$\m@th#1\widetilde{#2}$}%
	\ht\z@=.5\ht\z@
	\widetilde{\box\z@}}
\makeatother

\newtheorem{lemma}{Lemma}

\newtheorem{remark}{Remark}

\begin{document}
\newcommand{\pp}[1]{\textcolor{red}{#1}}

\title{Value-Based Reinforcement Learning for \\ Digital Twins  in Cloud Computing}

\author{ Van-Phuc Bui, \textit{IEEE Student Member}, Shashi Raj Pandey,\textit{ IEEE Member}, Pedro M. de Sant Ana, Petar Popovski, \textit{IEEE Fellow}\thanks{V.-P Bui, S.R. Pandey, and P. Popovski (emails: \{vpb, srp, fchi, petarp\}@es.aau.dk) are all with the Department of Electronic Systems, Aalborg University, Denmark. 
        P. M. de Sant Ana is with the Corporate Research, Robert Bosch GmbH, 71272 Renningen, Germany (email: Pedro.MaiadeSantAna@de.bosch.com). This work was supported by the Villum Investigator Grant ``WATER'' from the Velux Foundation, Denmark.}}

\maketitle
\thispagestyle{empty}
\begin{abstract}
    The setup considered in the paper consists of sensors in a Networked Control System that are used to build a digital twin (DT) model of the system dynamics. The focus is on control, scheduling, and resource allocation for sensory observation to ensure timely delivery to the DT model deployed in the cloud. Low latency and communication timeliness are instrumental in ensuring that the DT model can accurately estimate and predict system states.
    However, acquiring data for efficient state estimation and control computing poses a non-trivial problem given the limited network resources, partial state vector information, and measurement errors encountered at distributed sensors. 
    We propose the REinforcement learning and Variational Extended Kalman filter with Robust Belief (REVERB), which leverages a reinforcement learning solution combined with a Value of Information-based algorithm for performing optimal control and selecting the most informative sensors to satisfy the prediction accuracy of DT. Numerical results demonstrate that the DT platform can offer satisfactory performance while reducing the communication overhead up to five times.
\end{abstract}
\begin{IEEEkeywords}
    Digital twin, Reinforcement Learning, Internet of Things, Dynamic Systems, Sensor Networks.
\end{IEEEkeywords}

%%%%%%%%%%%%%%%%%%%%%%%%%%%%%%%%%%%%%%%%%%%%%%%%
\vspace*{-15pt}
\section{Introduction}\label{sec:intro}
%%%%%%%%%%%%%%%%%%%%%%%%%%%%%%%%%%%%%%%%%%%%%%%%
The Industry 4.0 smart manufacturing paradigm necessitates the acquisition of substantial volumes of real-time data, which emanate from a diverse array of wireless sensors \cite{tang2015tracking}. 
In contrast to conventional simulation tools or optimization methodologies, digital twin (DT) models undergo a transformation of these extensive datasets into predictive models. The utilization of these models allows for the emulation of potential control strategies, which in turn supports real-time interactions and decision-making for system operators~\cite{9899718}.

A network control system (NCS) representing the physical world, alongside a DT is situated either in the cloud, catering to extensive physical systems, or at the edge, tailored to local physical systems. The network comprises sensor devices and/or central/distributed units integral to a 5G system or beyond \cite{tse2005fundamentals}.
The acquired knowledge from the DT model serves twofold purposes: controlling the physical world; and providing monitoring and forecasting future states. Given the intricate interplay between communication and computation systems, devising a joint design strategy poses challenges in preserving the predictive efficacy of the DT model, and performing accurate control signals while simultaneously extending the operational lifespan of the network~\cite{ruah2023bayesian, 10092861}. There has been a substantial research on the use of Reinforcement Learning (RL) at the DT to perform various tasks~\cite{ruah2023bayesian}. 
In \cite{10092861}, the issue of scheduling IoT sensors is examined through the use of Value of Information (VoI) while taking into account the limitations of communication and reliability. The ultimate goal is to reduce the Mean Squared Error (MSE) of state estimation, which is often hindered by imprecise measurements. Additionally, authors in \cite{9768131} offers a possible resolution for scheduling sensing agents through the utilization of VoI, ultimately leading to enhanced accuracy in a variety of summary statistics for state estimation. 
These aforementioned works \cite{ruah2023bayesian,9768131, 10092861} and the reference therein, however, do not consider the influence of control performance in the physical world and the strategy of selecting effective sensing agents based on the reliability of estimates and latency requirements. %  9656153

\begin{figure}[t]
    \centering
    \includegraphics[trim=1.8cm 0.0cm 0.cm 0cm, clip=true, width=0.4\textwidth]{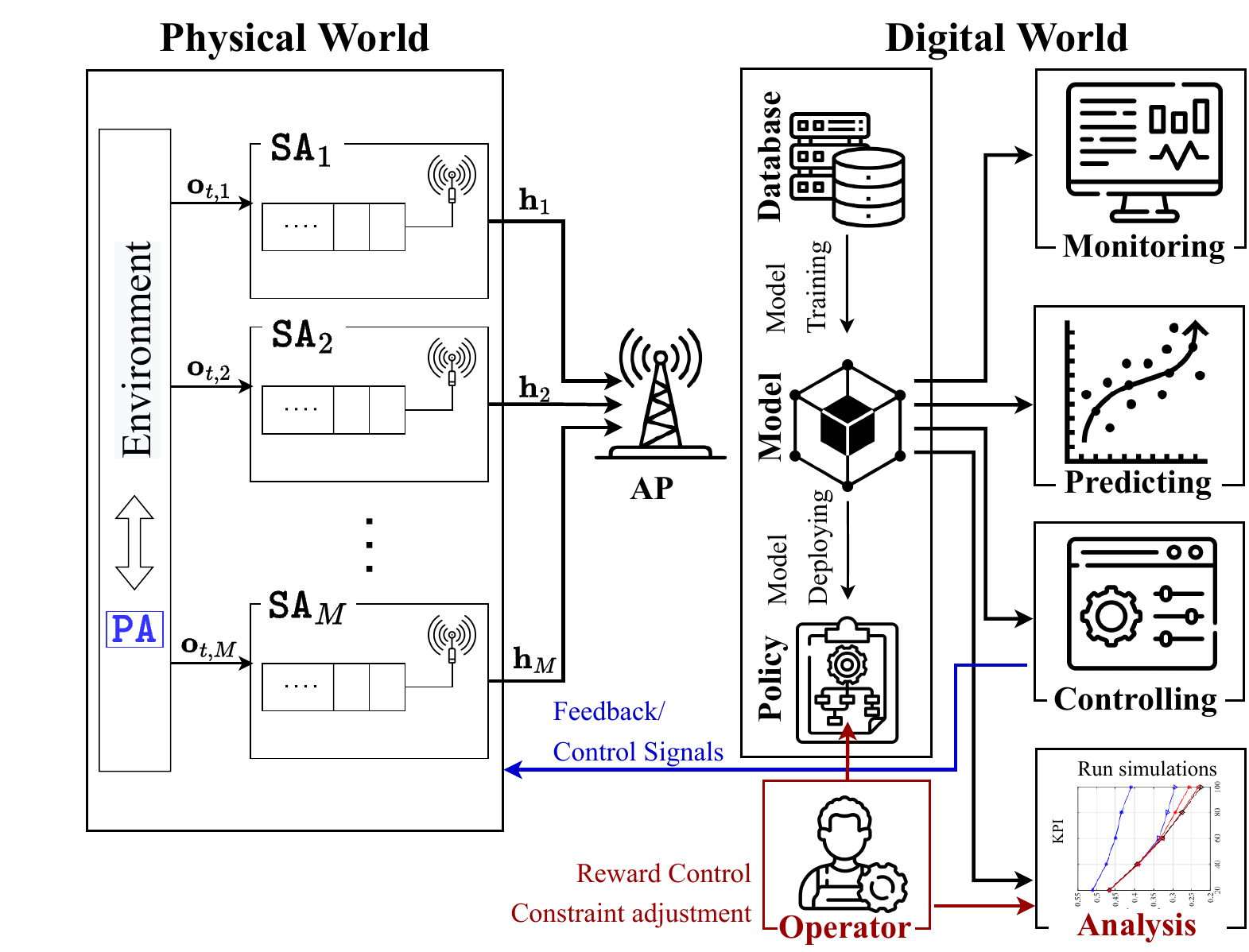}
    % \vspace*{-0.3cm}
    \caption{The considered architecture with a Digital Twin (DT).}
    \label{fig_system}
    \vspace*{-0.5cm}
\end{figure}

This paper addresses an optimization challenge that involves managing both the optimal control, accuracy of state estimation and power consumption in an NCS, illustrated in Fig.~\ref{fig_system}. Alongside this, a scheduling algorithm is introduced for $\SA$s, which takes into account the quality of their observations as well as the communication constraints of the DT model. Our contributions are listed as follows: $(i)$ we  introduce a DT architecture tracking dynamic changes of the system parameters and controlling system dynamics; $(ii)$ we propose an uncertainty control reinforcement learning framework that learn to perform actions while controlling the state' uncertainty estimation; $(iii)$ we formulate a novel optimization problem to efficiently schedule sensing agents for maintaining the confidence of DT's system estimate while minimizing the energy cost under latency requirements. We then propose a VoI-based algorithm resulting in a practical and efficient solution in polynomial time. $(iv)$ Numerical simulations were conducted to evaluate the algorithm's performance, which confirms that it surpasses other benchmarks in both controlling  and power consumption while improving DT estimation error.

%%%%%%%%%%%%%%%%%%%%%%%%%%%%%%%%%%%%%%%%%%%%%%%%
\vspace*{-5pt}
\section{Digital Twin Architecture}
%%%%%%%%%%%%%%%%%%%%%%%%%%%%%%%%%%%%%%%%%%%%%%%%
%	\vspace*{-10pt}
\subsection{System Model}

\begin{figure}[t]
    \centering
    \includegraphics[trim=0cm 0.0cm 0.cm 0cm, clip=true, width=0.5\textwidth]{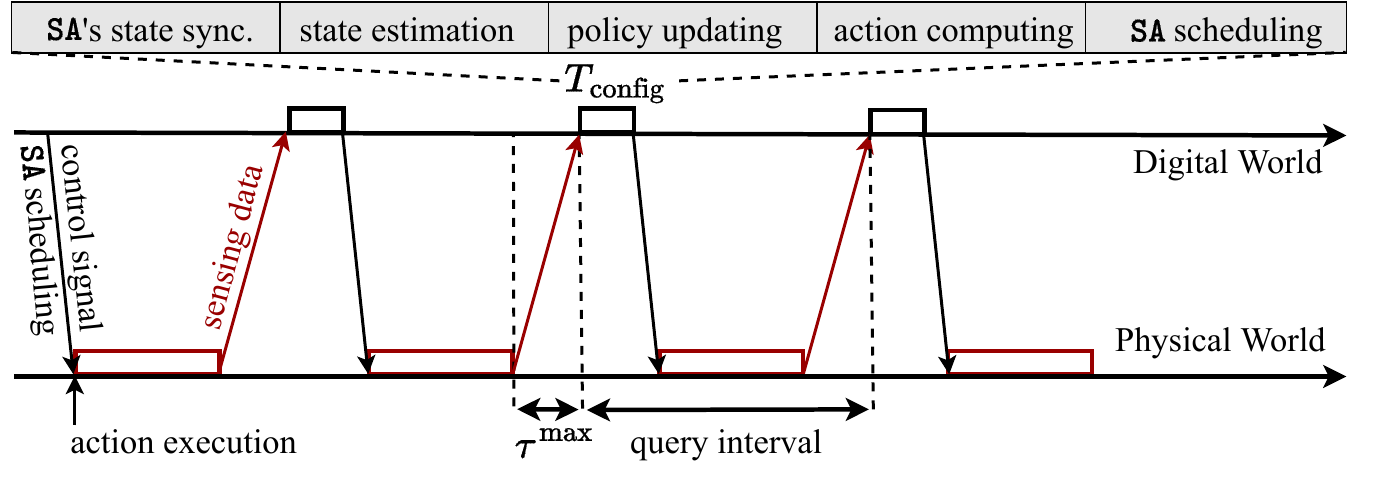}
    % \vspace*{-0.3cm}
    \caption{The communication diagram.}
    \label{fig_diagram}
    \vspace*{-0.7cm}
\end{figure}

We adopt a DT architecture, as illustrated in Fig.~\ref{fig_system}, including a single primary agent ($\PA$) and a set of sensing agents ($\SA$s) denoted by $\Mcal=\{1,2,\dots,M\}$. These $\SA$s are responsible for observing the environment and  communicate with the \emph{access point} (AP) through a  wireless channel, which facilitates the construction of the DT model for the $\PA$ and operates in Frequency Division Duplexing (FDD) mode. We assume the communication link between AP and the cloud is perfect. The communication diagram is illustrated in Fig.~\ref{fig_diagram}, where $T_\mathrm{config}$ accounts for the  configurable and computing time the DT.
At the beginning of each \emph{query interval} (QI), the DT model is required to update the active state of $\SA$s, after which it estimates the full system state, updates policy, computes optimal control signal, schedules at most $C$ $\SA$s and applies a fusion algorithm at the AP. Once a control command is generated, the controller promptly transmits it through a downlink channel to the physical world. The application output for actuators control, such as motor drives, retrieves the most recently stored command values from memory and applies them to drive the system dynamics. 

Each QI takes place at time instances $t \in \Tcal= \{1, 2, \dots, T\}$. To maintain synchronization and consistency, these $\mathtt{SA}$s periodically synchronize their DTs, including locations and power budget status, with the Cloud platform, ensuring a high degree of reliability and managing the overhead of synchronization. 
% Leveraging the current state of the DT model and the updated environmental information, the DT predicts the $\PA$'s state, devises an optimal policy for scheduling the $\SA$s, and generates control signal for subsequent actions. The feedback signal containing these insights is then conveyed to the AP, prompting it to execute appropriate actions in the physical world.
The $\PA$ engages in interactions with the environment, operating within a $K$-dimensional process as $\Kcal=\{1,2,\dots,K\}$. The state at the $t$-th QI is $\mathbf{s}_t= [s_t^1, s_t^2, \dots, s_t^K]^T$, and its evolution is described as:
\begin{align}\label{dynamic_model}
    \mathbf{s}_t &= f(\mathbf{s}_{t-1}) + \bB \ba_{t-1} + \mathbf{u}_t, \quad \forall t\in\Tcal, 
\end{align}
Here, $f:\mathbb{R}^{K}\rightarrow\mathbb{R}^K$ denotes the state update function, $\ba_{t-1}$ is the control signal, and the matrix $\bB$ describes the how the control impacts the dynamics. $\mathbf{u}_t\sim \mathcal{N}(\mathbf{0},\mathbf{C}_{\mathbf{u}})$ is the process noise.
At QI $t$, a $D$-dimensional observation  is received at  $\SA_m$  $(m \in \mathcal{M})$ as $\mathbf{o}_{t,m} = g(\mathbf{s}_t)+ \mathbf{w}_{t,m} \in\mathbb{R}^{D} (D\leq K)$  corresponding to the $\PA$'s state. To simplify the analysis, the observation $\bo_{t,m}$ is assumed to be linearly dependent on the system state, which is
% \begin{align}
    $\bo_{t,m}= \bH_m\bs_t+ \bw_{t,m}, \forall m\in\Mcal,$  with the observation matrix $\bH_m\in\mathbb{R}^{D\times K}$  and the measurement noise $\bw_{t,m}\sim \Ncal(\mathbf{0}, \mathbf{C}_{\bw_m})$.
    % \end{align}
%	where $\bH_m\in\mathbb{R}^{D\times K}$ indicates  the observation matrix, and $\bw_{t,m}\sim \Ncal(\mathbf{0}, \mathbf{C}_{\bw_m})$ is the measurement noise. 
The covariance matrices $\mathbf{C}_{\bu}$ and $\mathbf{C}_{\bw_m}$ are generally not diagonal. Let $\tau^\mathrm{max}$ be the maximum tolerable latency for transmitting $\SA_m$'s information, the system fulfills the application reliability at a QI $t$ if
\begin{align}\label{aoi_requirement}
    \mathbb{P}[\tau _{t,m} >  \tau^\mathrm{max}] \leq \varepsilon, \forall m\in\Mcal, t\in\Tcal,
\end{align}
with $\varepsilon$ is a outage probability parameter depending on system characteristics \cite{8017572}.

%	\vspace*{-0.4cm}
\subsection{Problem Formulation}
%	\vspace*{-0.4cm}

The DT model's objective is to uphold a precise estimate of  $\PA$'s state and offer the optimal sequence of actions to be executed in the physical realm based on its beliefs about states. 
%	A fundamental distinction within our system lies in its thorough consideration and evaluation of real-world environments, which often involve state observations that are characterized by noise or corresponded costs. 
Herein, the predicted estimator $\hat{\bs}_t$ of ${\bs}_t$ is modeled with 
%	\begin{align}
    %		p(\bs_t)\sim \mathcal{N}(\hat{\bs}_t, \bPsi_t), n \in\Ncal.
    %	\end{align}
$p(\bs_t)\sim \mathcal{N}(\hat{\bs}_t, \bPsi_t), n \in\Ncal$.  
The MSE of the estimator is
\begin{align}
    \text{MSE}_{} = \mathbb{E}\big[||\bs_t-\hat{\bs}_t||^2_2\big], t\in\Tcal. 
\end{align}

\begin{remark}\label{certainty}
    We define the maximum acceptable standard deviation for feature $k\in\Kcal$ as $\xi_k$. This corresponds to the following condition:
    \begin{equation}\label{qos_condition}
        \sqrt{[\bPsi_t]_k} \leq {\xi}_k, \forall k \in\Kcal,
    \end{equation}
    where $[\bPsi_t]_{k}$ is the $k$-th element of the diagonal of $\bPsi_t$.
\end{remark}
Defining  $V^\pi(\bs_0)$ as value function of controlling $\PA$ in \eqref{dynamic_model} under control policy $\pi$ and $\mathbf{p}^\mathrm{tx}_t$ as the transmitted power consumption to forward observations $\bo_{t} \triangleq \{\bo_{t,m}\}$ from $\SA$s to AP at QI $t$, we interested in joint minimizing the sum power consumption and delivering optimal control signals. We introduce the ultimate goal $h(\{\mathbf{p}^\mathrm{tx}_t\}, \{\ba_t\}) = [V^\pi(\hat{\bs}_0), -\sum_{t=0}^{\infty}\mathbf{p}^\mathrm{tx}_t]^T$, then formulate the  optimization problem  as
% Denoting $h\big(V^\pi(\bs_0), \sum_{t=0}^{\infty}\mathbf{p}^\mathrm{tx}_t\big)$ as the joint function of value function $V^\pi(\bs_0)$ of controlling under policy $\pi$ and transmitted power consumption $\{\mathbf{p}^\textbf{tx}_t\}$, we consider the optimization problem as
\begin{subequations} \label{globe_prob}
    \begin{alignat}{2}
        &  \underset{\{\mathbf{p}^\mathrm{tx}_t\}, \{\ba_t\}}  { \mathrm{maximize}} \ & &h\big(\{\mathbf{p}^\mathrm{tx}_t\}, \{\ba_t\}\big) \label{}\\
        &\mathrm{subject\ to } \ && \eqref{dynamic_model}, \eqref{aoi_requirement}, \eqref{qos_condition}.
    \end{alignat}
\end{subequations}
It is noted that we consider a scenario where the belief vector can be enhanced through estimation techniques facilitated by DT prior to being utilized by RL to suggest the optimal action as an output. By employing this approach, the accuracy of noisy observations can be enhanced, enabling the agent to make more precise decisions.  

In the following, we propose the REVERB (REinforcement learning and Variational Extended Kalman filter with Robust Belief) framework including two-step approach to address the problem \eqref{globe_prob}: $(i)$ we employ an uncertainty control RL algorithm to devise control actions for the physical world and effectively managing the state estimation errors; $(ii)$ the VoI-based $\SA$ scheduling with optimal power control algorithm is utilized to identify the most significant  $\SA$s for observing its sensing signals, guided by the requirements from the RL model and DT. To further enhance the accuracy of estimated states and forecast the forthcoming system state,  the Extended Kalman Filter (EKF) technique is revised and integrated. 

%%%%%%%%%%%%%%%%%%%%%%%%%%%%%%%%%%%%%%%%%%%%%%%%
        \vspace*{-0.3cm}
\section{Uncertainty Control POMDP}
%%%%%%%%%%%%%%%%%%%%%%%%%%%%%%%%%%%%%%%%%%%%%%%%
The control problem in \eqref{globe_prob} is considered as Partially Observable Markov Decision Process (POMDP)  that expands upon the MDP by incorporating the sets of observations and observation probabilities to actual states because the provided observations  only offer partial and potentially inaccurate information. 
In particular, a POMDP is presented by a 7-tuple $ \langle \Scal, \Acal, \Ocal, \mathtt{P}, \mathtt{O}, r ,\gamma \rangle$. In particular, $ \Scal $ is a the finite set of possible states, $ \Acal $ is a  set of control primitives and $ \Ocal $ denotes a  set of possible observations. At a time instant $ t $, the agent makes an action $\ba_t$ to move from state $ \bs_t $ to $ \bs_{t+1} $ with the transition probability  $ \Ptt = \mathbb{P}[\bs_{t+1}|\bs_{t},\ba_t]$. An observation $ \bo_{t+1} $ received from $\SA$s tracking system's state occurred with a probability $ \Ott = \mathbb{P}[\bo_{t+1}|\bs_{t+1},\ba_t] $. Also, as soon as the transition, the agent receives a numerical reward $ r(\bs_{t}, \ba_t, \bs_{t+1}) $ verifying	$ r(\bs_{t},\ba_t,\bo_{t+1}) \leq r^\mathrm{max} $. 
An agent do not know exactly its state at QI $t$ and it maintains a estimate-vector $\hat{\bs}_t$ describing the probability of being in a particular state $ \bs_t \in\Xcal $. 
We define $ \pi $ as a policy of the agent that specifies an action $ \ba_t $ based on its policy  $\pi(\hat{\bs}, \ba)$. 
With initial belief $\hat{\bs}_0$, the expected future discounted reward for policy $\pi(\hat{\bs},\ba)$ is given as $V^\pi(\hat{\bs}_0) = \mathbb{E}\left[\sum_{t=0}^{\infty}\gamma_t r(\bs_{t},\ba_t,\bs_{t+1}) |\hat{\bs}_0,\pi\right],$ where $0< \gamma_t <1 $ is the discount factor.

At QI $t$, the estimated state vector is  $\hat{\bs}_t  = [\hat{\bs}_t^1, \hat{\bs}_t^2, \dots, \hat{\bs}_t^K]^T  \in \mathbb{R}^K$, which is governed by $\hat{\bs}_t \sim p(\hat{\bs}_t|\bs_t; \boldsymbol{\eta}_t)$ as the  conditional probability distribution function (pdf) of $\hat{\bs}_t$ given $\bs_{t}$. $ \boldsymbol{\eta}_t =[ {\eta}_{t,1},  {\eta}_{t,2}, \dots, {\eta}_{t,K}]^T \in \mathbb{R}^K$ is parameterized by the accuracy vector, whose each element ${\eta}_{t,k}$ indicates the accuracy of $k$-th process of estimated state $\hat{\bs}_t$. In this work, we define $	{\eta}_{t,k} = \frac{1}{[\bPsi_t]_{k}} , \forall k\in \Kcal$. 
%	\begin{align}\label{eta_compute}
    %		{\eta}_{t,k} = \frac{1}{[\bPsi_t]_{k}} , \forall k\in \Kcal.
    %	\end{align}
%$\boldsymbol{\eta}_t^k = 1/ [\bPsi_t]_{k}$. 
As  ${\eta}_{t,k} $ increases, the confidence level of $\hat{s}_{t,k}$ also increases, facilitating the RL agent in making accurate decisions. Nevertheless, achieving high reliability of $\hat{s}_{t,k}$ necessitates low error of measurement $\SA$ or multiple observations, consequently amplifying both communication and processing expenses.
Given $\bs_{t}$ and $\boldsymbol{\eta}_t$, the $\boldsymbol{\hat{s}}_{t,k}$ is assumed exhibit statistical independence, meaning that we can express  
%	\begin{align}
    $p(\hat{\bs}_t|\bs_t; \boldsymbol{\eta}_t) = \prod_{k\in \Kcal} p(\boldsymbol{\hat{s}}_{t,k}|\bs_t; \boldsymbol{\eta}_t)$  
    %	\end{align}
in terms of their factorization.

%		\vspace*{-0.1cm}
\subsubsection{Actor-critic DRL Algorithm}
For training the policy $\pi(\boldsymbol{\hat{s}},\boldsymbol{a})$, we employ  Proximal Policy Optimization (PPO) with the actor-critic structure at RL agents that involves dividing the model into two distinct components, thus harnessing the strengths of both value-based and policy-based methods \cite{NIPS1999_6449f44a}. 
Specifically, the actor is primarily responsible for estimating the policy, which dictates the agent's actions in a given state, and the critic is dedicated to estimating the value function  predicting the expected future reward for a particular state or state-action pair. The actor's policy undergoes refinement based on the feedback provided by the critic. 
To address the joint design problem with the objective of optimizing actions while minimizing communication energy, we implement the Deep RL (DRL) approach within the DT cloud environment, where  the  action and reward are needed to be redefined.
%\subsubsection{State Formulation}
\subsubsection{Action Space Reformulation}
We focus on the issue of optimizing the accuracy of the estimated state by the RL agent, enabling it to select $\eta_{t,k}$ on a continuous scale. The underlying motivation for the proposed framework is to unveil the inherent characteristics of the observation space in terms of the informational value that the observations offer for the given task. 
The formulation of the action vector structure within the RL agent implemented in DT is represented as
\begin{align}
    \ba_t  = [{a}_{t,1},\dots,{a}_{t,Z},\eta_{t,1},\dots,\eta_{t,K}] \in\mathbb{R}^{ZK},
\end{align}
where $\mathcal{Z} (|\mathcal{Z}|= Z)$ is the action space and $\{{a}_{t,z}\}_{z\in\mathcal{Z}}$ correspond to the control signals that exert an influence on the physical environment, enabling the agent to advance towards its objective. Additionally,  $\eta_{t,k}\in[0,\infty]$ denotes the accuracy selection pertaining to the estimated state.

%	\vspace*{-0.1cm}
\subsubsection{Reward Function Reformulation}
It is imperative for the RL agent to not only navigate towards the primary objective defined for the problem but also acquire the ability to regulate the acceptable level of accuracy  $\{\eta_{t,k}\}_{k\in\Kcal}$. Consequently, the goal-based reward $r_t$ is transformed into an uncertainty-based reward as $\tilde{r}_t = f(r_t, \boldsymbol{\eta}_t),$ 
wherein $f(\cdot)$ is a monotonically non-decreasing function of $r_t$ and $\boldsymbol{\eta}_t$. 
In scenarios where a direct cost function, denoted as $c_k(\cdot)$, exhibits an upward trend with the accuracy of the observation $o_{t,k}$, a suitable additive formulation can be employed. Specifically, the modified reward, $\tilde{r}_t$, can be expressed by
\begin{align}\label{reward_shaping}
    \tilde{r}_t = r_t + \kappa  \sum_{k=1}^K c_k(\eta_{t,k}).
\end{align}
Here, $c_k(\eta_{t,k})$ represents a non-increasing function of $\eta_{t,k}$, and $\kappa \geq 0$ serves as a weighting parameter. Therefore, the primary objective of the agent is two-fold: to maximize the original reward while simultaneously minimizing the cost associated with the observations.

%%%%%%%%%%%%%%%%%%%%%%%%%%%%%%%%%%%%%%%%%%%%%%%%
\vspace*{-0.1cm}
\section{VoI-based $\SA$s Scheduling and Power Control}
%%%%%%%%%%%%%%%%%%%%%%%%%%%%%%%%%%%%%%%%%%%%%%%%
In this section, our focus is to involve the scheduling of $\SA$s based on three factors: $(i)$\underline{} the acceptable level of accuracy $\boldsymbol{\eta}_t$ for the estimated state, as determined by the RL agent;  $(ii)$ the requirement pertaining  to the accuracy  of the DT model as in \eqref{qos_condition}; and $(iii)$ the communication resources, determined by the system capacity and latency requirements \eqref{aoi_requirement}.

\vspace{-10pt}
\subsection{Sensing Agent Scheduling Problem }
We formulate a combined $\SA$s scheduling and power control problem, where our objective is to determine the $\SA$s that should engage in transmission. We introduce arbitrary variables $\bar{\xi}_{t,k}^2$  indicates desired DT's error level of system's state at QI $t$.  Given the reliability for the DT in \eqref{qos_condition} and the requisite level of accuracy $\boldsymbol{\eta}_t$ to uphold the precision of the RL model, the DT should meet the error constraints at QI $t$ as
\begin{equation}\label{glob_qos}
    [\bPsi_t]_k \leq \bar{\xi}_{t,k}^2 \triangleq \min\Big\{\xi_k^2, \frac{1}{\eta_{t,k}}\Big\}, \forall t\in\Tcal, k\in\Kcal.
\end{equation}
Initially, we establish the available $\SA$ set $\Pcal_t$  and the scheduling set $\Qcal_t$ at QI $t$  as $ \Pcal_t = \Mcal,\ 
\Qcal_t=\emptyset$. Denoting the power allocation vector $\mathbf{p}^\mathrm{tx}_t = \{p^\mathrm{tx}_{t,m}\}_{m\in\Mcal}$ as the variable, we formulate the optimization problem with specifications as
\begin{subequations} \label{scheduling_problem}
    \begin{alignat}{2}
        \mathbf{p}_t^{\mathrm{tx}*} &= \  \underset{\mathbf{p}^\mathrm{tx}_t}  { \mathrm{argmin}} \ & &(1-\alpha) \sum_{k\in\Kcal}\max\left\{\frac{[\bPsi(n)]_{k }}{\bar{\xi}^2_{k}}-1,0\right\} \nonumber\\
        &&&+\alpha \sum_{\mathclap{m\in\Pcal(n)}}p^\mathrm{tx}_m(n)\label{scheduling_problema}\\
        &\mathrm{subject\ to }\  && |\Qcal_t| \leq C, \label{scheduling_problemb}\\
        &&& \mathbb{P}[\tau_{t,m} > \tau^\mathrm{max}] \leq \varepsilon, \forall m\in\Mcal,  \label{scheduling_problemc}
    \end{alignat}
\end{subequations}
wherein the non-negative parameter $\alpha\in[0,1]$ represents the relative weight to accuracy and energy efficiency within the underlying objective function. It is observed that objective \eqref{scheduling_problema} represents a relaxation of constraint \eqref{glob_qos} due to its dependence on practical conditions, i.e., in situations where the error surpasses a certain threshold, even  querying all sensor data fails to guarantee the desired level of reliability $\bar{\xi}^2_{t,k}$.
We note that obtaining observations from additional sources $\SA$s leads to an enhancement in estimation accuracy; however, this improvement comes at the cost of compromising energy efficiency. For those particular $\SA$s that exhibit considerable errors in their measurements or possess features that do not significantly contribute to meeting the confidence requirements of the $\PA$ (i.e., those with a low VoI), measuring and transmitting observations leads to an unnecessary expenditure of energy. The constraints given by \eqref{scheduling_problemb} and \eqref{scheduling_problemc} are required in ensuring communication capacity and the reliable execution of latency satisfaction within a FDMA uplink slot.  It is important to note that the optimization problem presented as \eqref{scheduling_problem} is inherently non-convex due to the non-convex nature of the objective function \eqref{scheduling_problema}, as well as the the constraints \eqref{scheduling_problemb}, \eqref{scheduling_problemc}. Furthermore, the node selection aspect renders the problem analogous to the classic NP-hard knapsack problem. To derive an efficient suboptimal solution, a heuristic algorithm based on EKF is employed.

\vspace*{-0.1cm}
\subsection{Sensing Agent Selection with Extended Kalman Filter}
Due to the complexity of sensor selection based on VoI, we adopt a heuristic approach with primary concept guiding the resolution of \eqref{glob_qos} is to ensure that, during each QI $t$ the minimum necessary number of $\SA$s is selected for transmission. This selection aims to sustain the desired level of certainty in estimating the state $\bs_t$ while saving communication resource for other purposes.
The expression for the Minimum Mean Square Error (MMSE) estimator applied to a KF  is provided in \cite[Eq. (1)]{Kalman}. In this work, for aiming to minimize the (weighted) variance of state components, we employ the Extended Kalman estimator, which is common in the IoT literature~\cite{huang2019epkf}. 
We also assume that the virtual environment possessing complete awareness regarding the process statistics, encompassing the update function $f(\mathbf{s})$ as well as the noise covariance matrices. The primary strategy to solve \eqref{scheduling_problem} involves selecting the minimum number of $\SA$s to transmit at each QI $t$ in order to maintain the required level of estimation certainty for state $\bs_t$ as specified by $\{\bar{\xi}_{t,k}^2\}$. Our proposed algorithm are summarized in Algorithm~\ref{scheduling_alg}, effectively addresses problem \eqref{scheduling_problem}.

In the dynamic setup, the initial state $\bs_t$ is considered a random vector characterized by a specific mean $\mathbb{E}[\bs_t] = \boldsymbol{\mu}_{\bs_0}$ and covariance matrix $\text{Cov}[\bs_0]= \mathbf{C}_{\bs_0}$.  $\Qcal_t$ is initialized as an empty set due to the absence of any prior information. The  EKF then calculates the estimation errors for the belief 
\begin{align}
    \hat{\bs}_t \sim \mathcal{CN}(\bmu_{\hat{\bs}_t},\bPsi^{\mathtt{pr}}_t)
\end{align}
at the $\PA$ based on prior updates $\hat{\bs}_{t-1}$ as described by 
\begin{align}\label{prior_error}
    \bPsi^{\mathtt{pr}}_{\hat{\bs}_t} = \bP\bPsi_{\hat{\bs}_{t-1}}\bP^T + \bC_{\bu_t},
\end{align}
where the Jacobian matrix $\bP= \mathcal{J}\{f(\bs_{t-1})\}$  linearizes the nonlinear model of $f(\bs_{t-1})$. Subsequently, for a given set of error variance qualities $\{\bar{\xi}_{t,k}^2\}_{t\in\Tcal, k\in\Kcal}$, the conditions specified in \eqref{glob_qos} lead to two potential cases: $(i)$ If \eqref{glob_qos}  are satisfied for all $k\in\Kcal$, the DT model achieves  bounds without requiring $\SA$s' observations. The prior update  suffices to establish the necessary confidence in the estimate, resulting in an empty set for $\Qcal^*_t = \emptyset $. ${(ii)}$ If any of conditions is violated, at least one process feature lacks sufficient accuracy, the acquisition of the corresponding observations becomes essential to enhance the estimation process, as dictated by the scheduling approach implemented in our proposed heuristic.
%	\begin{enumerate}
    %		\item If \eqref{glob_qos}  are satisfied for all $k\in\Kcal$, the DT model achieves  bounds without requiring $\SA$s' observations. The prior update  suffices to establish the necessary confidence in the estimate, resulting in an empty set for $\Qcal^*_t = \emptyset $.
    %		\item If any of conditions is violated, at least one process feature lacks sufficient accuracy, the acquisition of the corresponding observations becomes essential to enhance the estimation process, as dictated by the scheduling approach implemented in our proposed heuristic.
    %	\end{enumerate}

In the first  scenario, the  the belief $\hat{\bs}_t$ can be achieved through the EKF blind update  as 
\begin{align}\label{mu_prior}
    \hat{\bs}_t = {\hat{\bs}^{\mathtt{pr}}_t}  =f(\hat{\bs}_{t-1}) +\bB \ba_{t-1} + \bu_t, \forall t\in\Tcal.
\end{align}
In the second case, Algorithm~\ref{scheduling_alg} is utilized to identify the $\SA$s with the highest VoI for querying their observations.
In order to identify the most suitable candidate feature $s_{t,k}^{*}$, where $k\in\Kcal$, an optimization problem is formulated at the $i$-th iteration as
\begin{subequations} \label{finding_state}
    \begin{alignat}{2}
        s_{t,k}^{*} =& \underset{s_{t,k}\in\bs_t}{\ \mathrm{argmax }} &\quad & 
        {[\bPsi^{(i)}_t]_{k }}/{\bar{\xi}^2_{t,k}} \\
        &\mbox{subject to} &&  \SA_{m} \in\mathcal{P}_t, \forall m\in\Mcal, \label{finding_stateb}\\
        &&& \SA_{m} \rightarrow s_{t,k}, \forall m\in\Mcal,
    \end{alignat}
\end{subequations}where the notation $\SA_{m} \rightarrow s_{t,k}$ signifies that the $\SA_m$ measures feature $s_{t,k}$. We note that  at the $i$-th iteration, if any constraint in \eqref{glob_qos} is still unmet and $	|\Qcal_t| <\ C,\ |\mathcal{P}_t| >0$, 
there is room for scheduling new $\SA$s to join $\Qcal_t$. 
According to constraint \eqref{finding_state}, feature $s_{t,k}^{*}$ is selected only if at least one $\SA_{{m}}\in \Pcal_t$ can provide coordinating observations. Then,  $\SA_m^*\in\mathcal{P}_t$ measuring feature $s_{t,k}^{*}$ with the minimum error covariance is chosen to send its measurement. The scheduled and available $\SA$ sets $\Qcal_t$ and $\mathcal{P}_t$ are updated as
\begin{equation}\label{update_set}
    \Qcal_t \leftarrow \Qcal_t \cup\{ \SA_m^* \};\  \Pcal_t \leftarrow\Pcal_t\backslash\{\SA_{{m}}^*\}.
    % \vspace{-0.1cm}
\end{equation}
$\bH_t$ and $\bC_{\bw_t}$ are the combination observation and covariance matrices, which are respectively formulated as
\begin{align}
    \bH_t &= [\bH_{1};\bH_2;\dots;\bH_{|\Qcal_t|}], \label{H_compute} \\
    \bC_{\bw_t} &=  \text{diag}[\bC_{\bw_1}, \bC_{\bw_2},\dots,\bC_{\bw_{|\Qcal_t|}}]\label{Cw_compute},
\end{align}
where $\bH_{{m}}$ is the  observation matrix of  the $\SA_m ({{m}}\in\Qcal_t)$. From here, we can compute the EFK gain by
\begin{align}
    \bK_t =  \bPsi^{\mathtt{pr}}_{\hat{\bs}_t}{\bH}_t^T\big(\bC_{\bw_t}  + {\bH}_t \bPsi^{\mathtt{pr}}_{\hat{\bs}_t}{\bH}_t^T\big)^{-1},
\end{align}
The posterior error covariance matrix is derived by
\begin{align}\label{pos_variance}
    \bPsi^{\mathtt{pos}}_{\hat{\bs}_t}=  (\bI - \bK_t\bH_t) \bPsi^{\mathtt{pr}}_{\hat{\bs}_t},
\end{align}
The iterative loop continues while all three conditions are true:
\begin{align}
    \left\{\begin{matrix*}[l]\label{checking3_conditions}
        \{|\Qcal^*_t| &< C;\\ 
        \exists\ [\bPsi^{\mathtt{pos}}_{\hat{\bs}_t}]_{k} &\geq \bar{\xi}^2_{k}; \\ 
        \exists\ s_{t,k}^{*} &\mbox{ as a solution in }\eqref{finding_state}\}.
    \end{matrix*}\right.
\end{align}
It makes intuitive see that the loop repeats for at maximum $C$ iterations before terminating. The posterior update is then
\begin{equation}\label{pos_update}
    {\hat{\bs}^{\mathtt{pos}}_t}  = {\hat{\bs}^{\mathtt{pr}}_t} + \bK_t(\bo_t - \bH_t{\hat{\bs}^{\mathtt{pr}}_t} ),
\end{equation}
where $\bo_t $ represents the combination of received $\SA$ observations. Accordingly, we update $\hat{\bs}_t = \hat{\bs}^{\mathtt{pos}}_t$. Our approach ensures long-term balance between state certainty and communication cost with respect to the $\PA$'s requirements, despite the local scheduling solution provided by $\SA$ for different QIs.
%	The proposed \textbf{Algorithm~\ref{scheduling_alg}} has a worst-case complexity of $\mathcal{O}((DM)^{3} + K^2)$, assuming  $DM \geq K$.
\subsection{Optimal Power Scheduling}
Both DL and UL transmissions are subject to potential latency during data delivery, which is influenced by the channel quality and the scheduled transmission resource. 
The channel between $\SA_m$ and AP is modeled as Rician channel with strong LoS link and small-scale fading with rich scattering, where the instantaneous Signal-to-Noise (SNR) ratio is modeled as
\begin{align}\label{SNR}
    \gamma_{t,m} = \frac{\Gamma p_{t,m}^\mathrm{tx}}{d^{\alpha}_mWN_0} \mathcal{G}_m, \forall m \in \Mcal,
\end{align}
with  $\Gamma$ is constant depending on the system parameter (operating frequency and antenna gain); $p_{t,m}^\mathrm{tx}$ is the transmitted power, $d_m$ represents the distance between device $m$ and AP. $\alpha$ is the path loss exponent; $W$ is the allocated bandwidth and $N_0$ stands for noise power. $\mathcal{G}_m$ represents the fading power with expected value $\bar{\mathcal{G}} \triangleq \mathbb{E}[|\mathcal{G}_m|^2] =1$. Herein, we adopt $\mathcal{G}_m$ as Rician distribution with a non-central
chi-square probability distribution function (PDF) which is expressed as \cite{simon2008digital}
\begin{align}
    f_\mathcal{G}(w) = \frac{(G+1)e^{-G}}{\bar{\mathcal{G}}}e^\frac{-(G+1)w}{\bar{\mathcal{G}}}I_0\Big(2\sqrt{\frac{G(G+1)w}{\bar{\mathcal{G}}}}\Big),
\end{align}
where $w\geq 0$ and $I_0(\cdot)$ denotes the zero-order modified Bessel function of the first kind, while $G$ represents the Rician factor, signifying the ratio between the power within the Line-of-Sight (LoS) component and the power distributed among the non-LoS multipath scatters. 
We use Shannon’s bound to achieve the channel capacity $R_{t,m}$ of every link, which is $R_{t,m} = W\log_2(1+\gamma_{t,m})$. The optimal transmission power  for each scheduled $\SA$ is computed through the following lemma.
\begin{lemma}\label{optimal_bandwidth}
    Given the Rician factor $G$, the specific transmission bandwidth $W$ and the distance $d_m$, the optimal allocated power $p^*_{t,m}$ of $\SA_m$ to satisfy \eqref{aoi_requirement} is lower bounded by
    \begin{equation}\label{power_require}
        p^*_{t,m} = \frac{2 WN_0(1+G)(2^{D/(\tau^{\mathrm{max}}W)}-1)}{y_Q^2d_m^{-\alpha}\Gamma},
    \end{equation}
    where $y_Q = \sqrt{2G}+\frac{1}{2Q^{-1}(\varepsilon)}\log(\frac{\sqrt{2G}}{\sqrt{2G}-Q^{-1}(\varepsilon)})-Q^{-1}(\varepsilon)$, $D$ is the length of data packet, and $Q^{-1}(\cdot)$ is the  inverse Q-function.
    \begin{proof}
        The outage probability \eqref{aoi_requirement} is  formulated as
        \begin{align}
            \mathbb{P}[\Delta _{t,m} >  \tau^\mathrm{max}] &=  \mathbb{P}[W\log_2(1+ \gamma_{t,m}) <  \frac{D}{\tau^\mathrm{max}}] \\
            &=  \mathbb{P}[ \gamma_{t,m} <  2^\frac{D}{W\tau^\mathrm{max}}-1] \\
            & \triangleq 1 - Q(x_\tau, y_\tau) \leq \varepsilon, \label{y_Q_ex}
        \end{align}
        where $x_\tau =\sqrt{2G}$ and
        \begin{align}\label{y_tau_1}
            y_\tau = \sqrt{2(G+1)(2^\frac{D}{W\tau^\mathrm{max}}-1)\mathcal{G}/\gamma_{t,m}}, 
        \end{align}
        with $Q(x_\tau, y_\tau)$ as the the first order Marcum Q-function. At the maximum tolerable value of $\varepsilon$, then $y_Q = Q^{-1}(x_Q, 1-\varepsilon)$ formed as the inverse Marcum Q–function respecting to second argument. In this study, we consider a Rician channel with strong LoS component, i.e., $G > G_0^2/2$ and $Q^{-1}(\varepsilon) \neq 0$,  which yields the approximated form of $y_\tau$ as \cite[Lemma 1]{8017572}: 
        $y_\tau = \sqrt{2G}+ \frac{1}{2Q^{-1}(\varepsilon)}\log\Big(\frac{\sqrt{2G}}{\sqrt{2G}-Q^{-1}(\varepsilon)}\Big) - Q^{-1}(\varepsilon)$, 
        with  $G_0$ is  the intersection of the sub-functions at $x_\tau > \max[0, Q^{-1}(\varepsilon)]$. Plug this into \eqref{y_tau_1}, we obtain \eqref{power_require} and complete the proof.
    \end{proof}
    %	\vspace{-10pt}
\end{lemma}
\begin{algorithm}[t]
    \begin{algorithmic}[1]{\fontsize{8.5pt}{8.5pt}\selectfont
            \protect\caption{$\SA$ scheduling algorithm for problem  \eqref{scheduling_problem}} % \eqref{probGlobal}}
        \label{scheduling_alg}
        \global\long\def\algorithmicrequire{\textbf{Input:}}
        \REQUIRE $\bb_{0,\bo_0}, \bmu_{\bu_0}, C_{\bu_0}$
        Available uplink slots $C$,  The state and requirement certainty $\big(\bs, \{\bar{\xi}_{k}^2\}\big)$
        \global\long\def\algorithmicrequire{\textbf{Output:}}
        \REQUIRE The scheduled user set $\{\Qcal^*_t\}$; their belief $\{\hat{\bs}^*_t, \bPsi^*_t\}$, and the associated transmit power $\{p^\mathrm{tx}_{t, m}\}$
        \STATE Initial $\mathcal{Q}_t = \emptyset$
        \STATE Compute the prior errors $\bPsi^{\mathtt{pr}}_t$ as in \eqref{prior_error} %based on  prior  estimation.
        \IF {$[\bPsi^{\mathtt{pr}}_t]_{k} \leq\bar{ \xi}_{k}^2, \forall k$}
        \STATE Compute ${\hat{\bs}^{\mathtt{pr}}_t} $ as in \eqref{mu_prior}
        \STATE Update ${\hat{\bs}_t}^*  = {\hat{\bs}^{\mathtt{pr}}_t} $ and $\bPsi_{\hat{\bs}_t}^* = \bPsi_{\hat{\bs}_t}^\mathtt{pr}$
        \ELSE
        \STATE Set $i=1$ 
        \WHILE{conditions \eqref{checking3_conditions} hold}
        \STATE Update $s^*_{t,k}$ by solving \eqref{finding_state}
        \STATE Update $\Qcal_t$ and $\Pcal_t$ as in \eqref{update_set}
        \STATE Update the $\bK_t$, $\bH_t$ and $\bC_{\bw_t}$ as in \eqref{H_compute}, and \eqref{Cw_compute}
        \STATE Compute $	\bPsi^{\mathtt{pos}}_{\hat{\bs}_t}$ as in \eqref{pos_variance}
        \STATE  Set $i=i+1$
        \ENDWHILE
        \STATE Update $\Qcal^*_t = \Qcal_t$
        \STATE Compute  $\bPsi_{\hat{\bs}_t} ^*= \bPsi_{\hat{\bs}_t}^\mathtt{pos}$ and ${\hat{\bs}_t}^*  = {\hat{\bs}^{\mathtt{pos}}_t} $ as in   \eqref{pos_variance} and \eqref{pos_update}
        \STATE Update $\boldsymbol{\eta}^*_t$, and power transmission  $	\mathbf{p}_t^{\mathrm{tx}*}  =  \{p^\mathrm{tx}_{t,m}\}$ as in \eqref{power_require}
        \ENDIF
    }
\end{algorithmic}
\end{algorithm}

%%%%%%%%%%%%%%%%%%%%%%%%%%%%%%%%%%%%%%%%%%%%%%%%
\vspace{-0.4cm}
\section{Numerical Results}
%%%%%%%%%%%%%%%%%%%%%%%%%%%%%%%%%%%%%%%%%%%%%%%%

\begin{table}[!t]
\vspace{-6pt}
\caption{Simulation Parameters}
\resizebox{8.9cm}{!} 
{
    \begin{tabular}{ll|ll}
        \hline
        Parameter & Value & Parameter & Value \\
        \hline
        Carrier frequency ($f_c$)          & 2.4 GHz      &  Maximum latency ($\tau^\mathrm{max}$)          &   5 ms    \\
        Bandwidth   (W)       & 5 MHz      &  Channel noise power          & -11.5 dBm     \\
        DT required error variances $(\xi^2_{pos}, \xi^2_{vel})$ & (0.01, 0.001)     &  Max. distance ($d^{\max}$)          & 20 m     \\
        Max. connection ($C$) & 10 &  Leaning rate (both actor and critic) & $1e^{-3}$ \\ 
        Rician factor ($G$) & 15 dB &  Outage probability factor $(\varepsilon)$&  $1e^{-5}$\\
        System parameter ($\Gamma$) & 1 &  Data packet size $(D)$&  $1024$ bits\\
        \hline\vspace{-1.2cm}
    \end{tabular}
}
\label{parameters}
\end{table}

\begin{figure}[t]
\begin{minipage}{0.15\textwidth}
    %		\centering
    \includegraphics[trim=0.cm 0cm 0.cm 0cm, clip=true, width=1.1in]{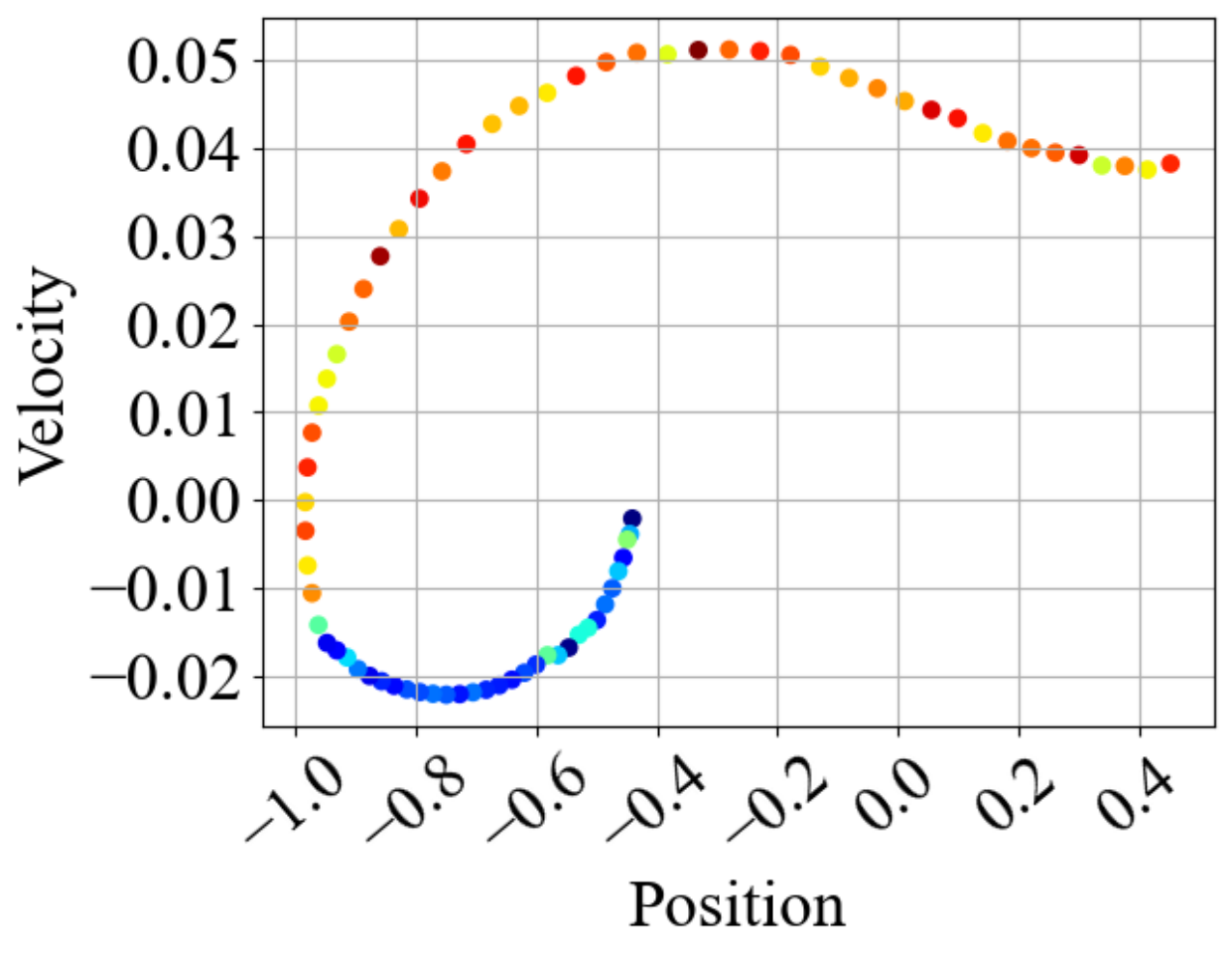} \\ 
    \vspace*{-0pt}
    \centering {\footnotesize$(a)$  REVERB \\(75 QIs)}
    \vspace*{-0pt}
\end{minipage}
\begin{minipage}{0.15\textwidth}
    %		\centering
    \includegraphics[trim=0.cm 0cm 0.cm 0cm, clip=true, width=1.1in]{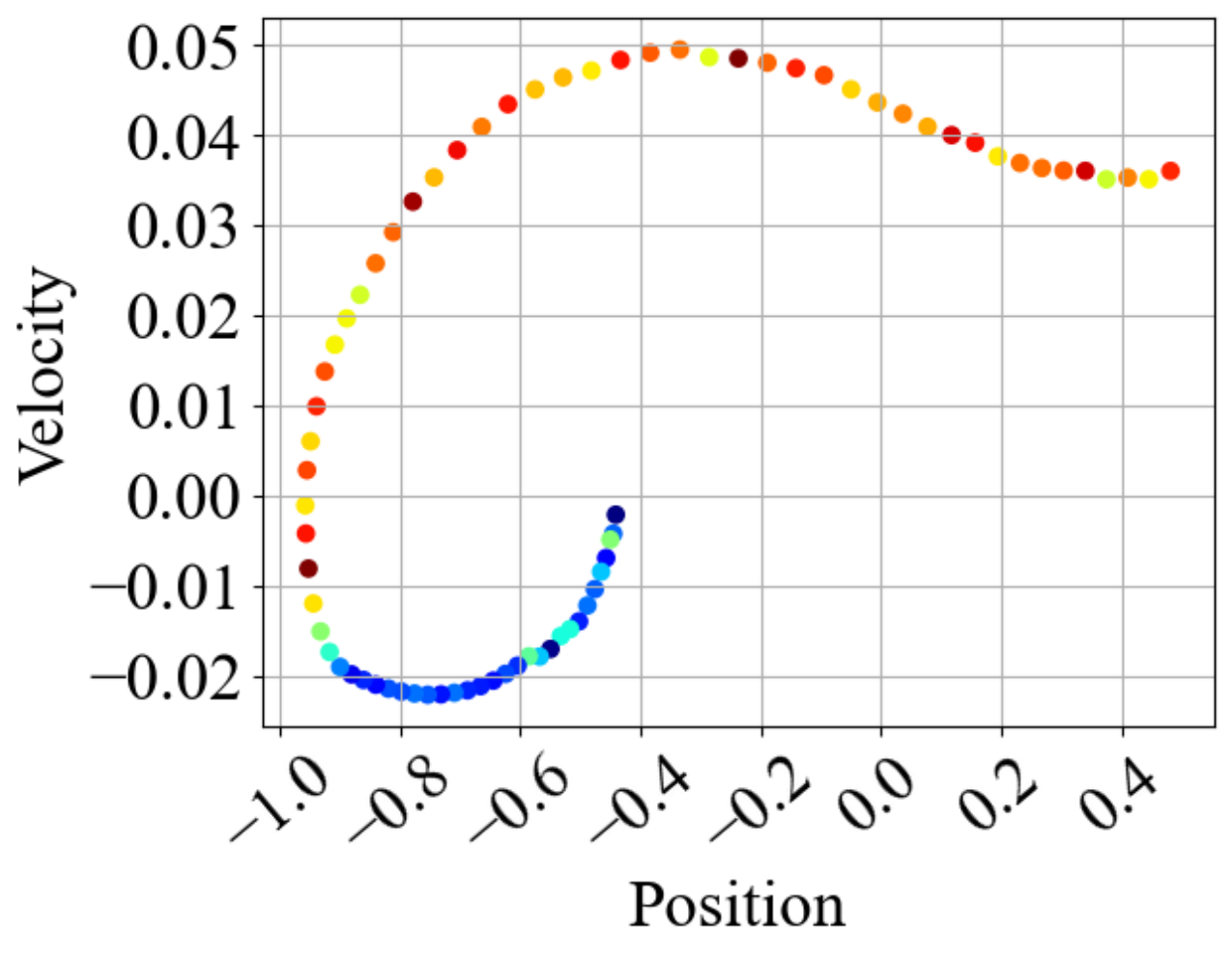} \\ 
    \vspace*{-0pt}
    \centering {\footnotesize$(b)$  Perfect \\ (74 QIs) }
    \vspace*{-0pt}
\end{minipage}
\begin{minipage}{0.16\textwidth}
    %		\centering
    \includegraphics[trim=0cm 0cm 0.cm 0cm, clip=true, width=1.15in]{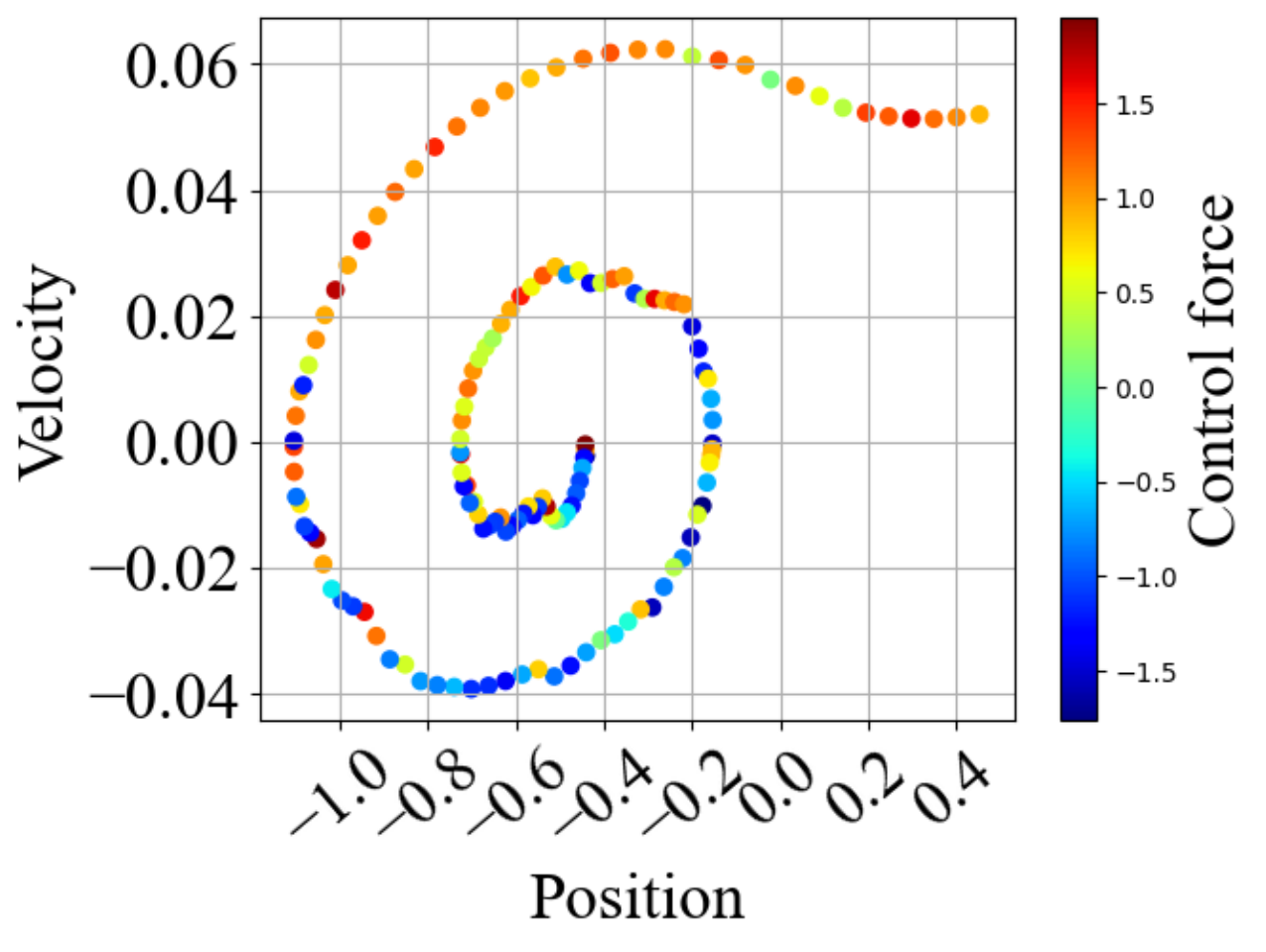} \\ 
    \vspace*{-0pt}
    \centering {\footnotesize$(c)$  Traditional \\ (143 QIs)}
    \vspace*{-0pt}
\end{minipage}
\begin{minipage}{0.165\textwidth}
    %		\centering
    \includegraphics[trim=0.cm 0cm 0.cm 0cm, clip=true, width=1.15in]{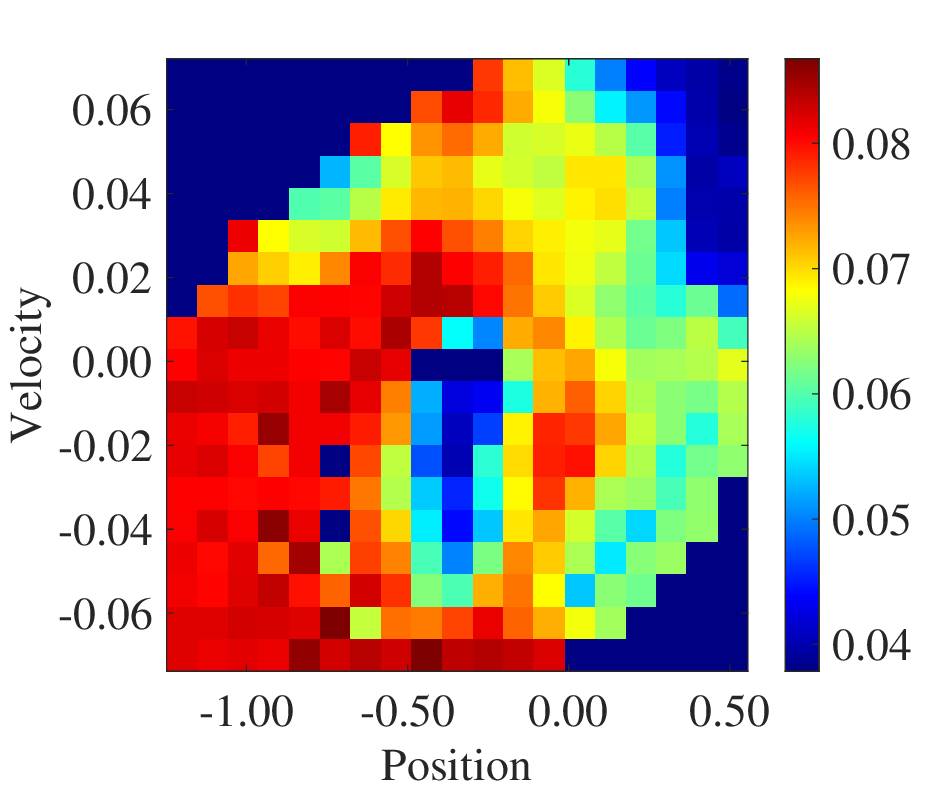} \\ 
    \vspace*{-5pt}
    \centering {\footnotesize$(d)$  Position uncertainty}
    \vspace*{-0pt}
\end{minipage}
\begin{minipage}{0.16\textwidth}
    %		\centering
    \includegraphics[trim=0.cm 0cm 0.cm 0cm, clip=true, width=1.15in]{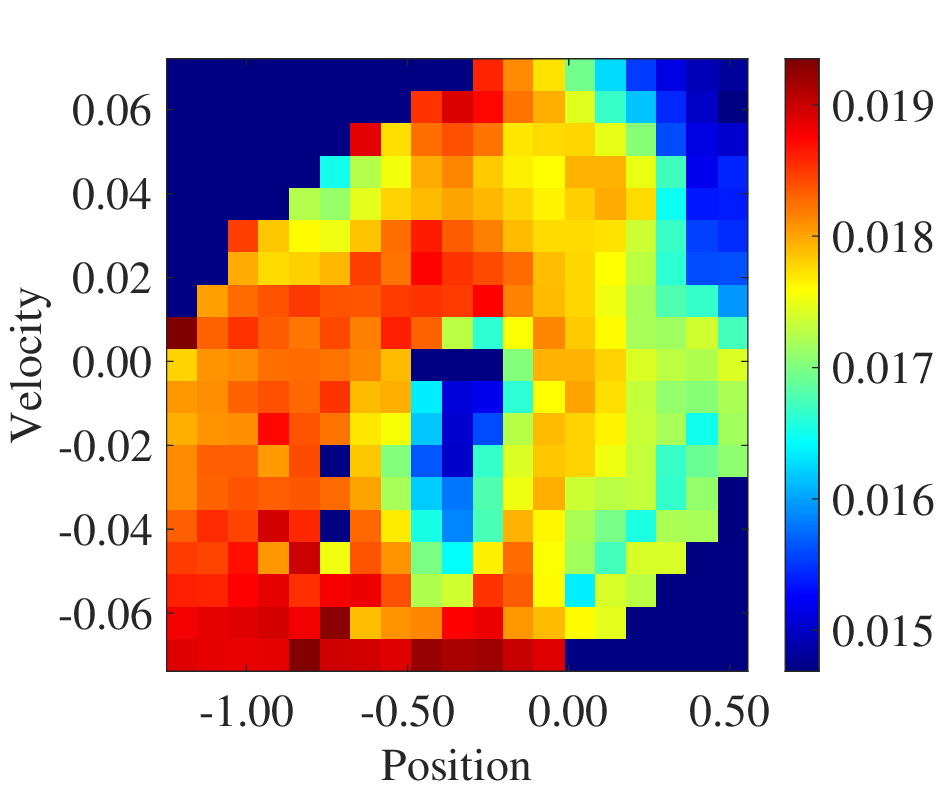} \\ 
    \vspace*{-5pt}
    \centering {\footnotesize$(e)$  Velocity uncertainty}
    \vspace*{-10pt}
\end{minipage}
\begin{minipage}{0.15\textwidth}
    %		\centering
    \includegraphics[trim=0cm 0cm 0.cm 0cm, clip=true, width=1.15in]{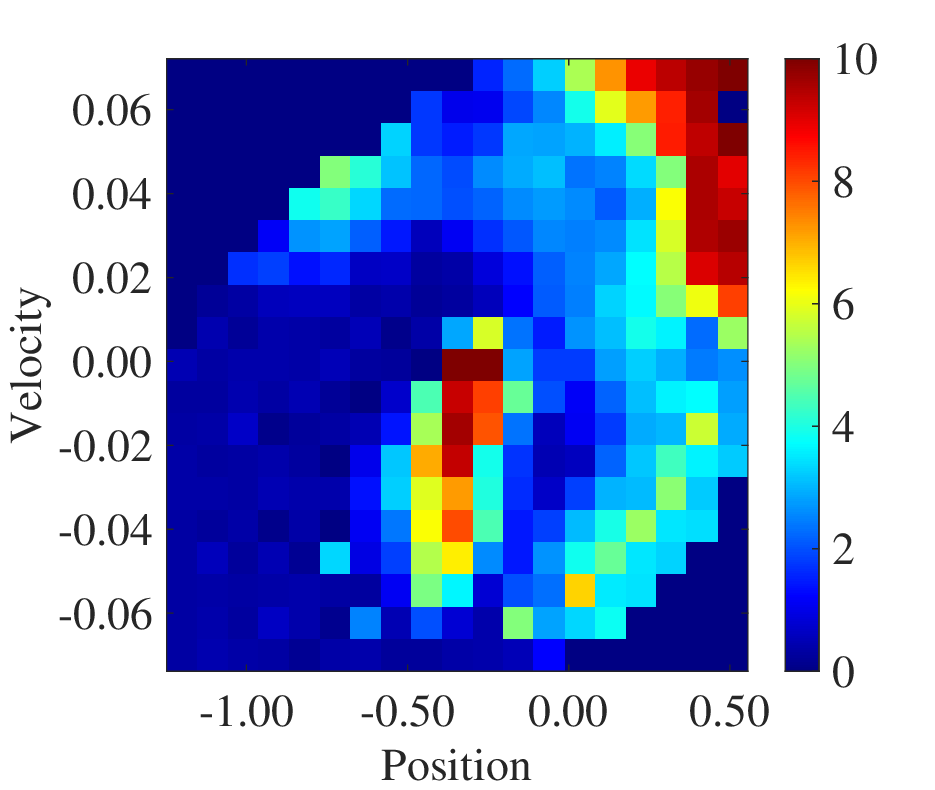} \\ 
    \vspace*{-5pt}
    \centering {\footnotesize$(f)$  No. selected $\SA$s}
    \vspace*{-0pt}
\end{minipage}
\caption{Performance evaluation: $(a)$, $(b)$ and $(c)$ compare the trajectory with coordinated force of different schemes; $(d)$ and $(e)$ are position and velocity uncertainty level ($\{\bar{\xi}_{t,k}\}$) of REVERB; and $(f)$ represents  number of selected $\SA$s under position-velocity coordinate.}
\label{fig_trajectory}
\vspace*{-15pt}
\end{figure}

We examine our proposed REVERB algorithm with the MountainCarContinuousv0 environment from the OpenAI Gym \cite{moore1990efficient}. The state vector $\bs_{t} = [x_t, \dot{x_t}]^T$ includes position $x_t$ and velocity $\dot{x_t}$. The observation matrices are given by $\bH_{pos} = [1,  0]; \text{ and } \bH_{vel} = [0,  1]$,  
respectively for the position and velocity. Other important parameters are included in Table~\ref{parameters}. In the DT system, the agent makes decisions concerning the applied force $a_t\in[-1,1]$ on the car and selects an accuracy level denoted by $\boldsymbol{\eta}_t = [\eta_{t,1}, \eta_{t,2}]^T$. The original reward for each QI in the environment is  ${r}_t = -0.1 \times a_t^2.$
%\begin{align}
%{r}_t = -0.1 \times a_t^2.
%\end{align}
In our system, we adopt a modified reward indicated in \eqref{reward_shaping} as
\begin{align}\label{reward_shaping1}
\hat{r}_t = {r}_t + \kappa \times \Big(\frac{1}{2}\sum_{i =1}^{2}\eta_{t,i}\Big),
\end{align}
where the positive weighted parameter $\kappa = 5\times 10^{-6}$. Additionally, the original environment includes a termination reward when the car successfully reaches the target position at 0.45. The evolution of the state in \eqref{dynamic_model} is  defined with
\begin{align}
f(\bs_{t-1})&= \begin{bmatrix}
    \dot{x}_{t-1}	\\ 
    -\varphi \cos(3x_{t-1})	
\end{bmatrix}, \quad 
\bB &= [0, \vartheta]^T,
\end{align}
where the constants $\varphi = 0.0025$ and $\vartheta = 0.0015$. $M$ $\SA$s are assuming placed randomly within an area where the maximum distance $d^\mathrm{max}$ from $\SA$s to AP.

We compare our proposed REVERB and four other benchmarks: \textit{Perfect} allows DT to get the observation of the next state without any error; \textit{Cost-based} greedy indicates that the AP queries all nearest $\SA$s based on ascending order of distance from AP to $\SA$ at each QI. \textit{Error-based }greedy \cite{10092861} is similar to \textit{Cost-based} but it relies on the decreasing confidence levels of $\SA$s. In the greedy benchmarks is $|\Qcal_t^*| = C$. We also consider the \textit{Traditional} scheme~\cite{sutton2018reinforcement} where the RL agent gets noise observation from one $\SA$ every QI without Algorithm 1. All schemes are conducted under 1000 Monte Carlo simulations.

Figure~\ref{fig_trajectory}(a), (b) and (c) compare the trajectory evolution with coordinated force of REVERB, \textit{Perfect}, and \textit{Traditional} schemes. It is observed that REVERB's trajectory is close to the \textit{Perfect} when using only 1 step more to reach the goal. On the contrary, the trained network with the \textit{Traditional}  must use up to $2\times$ number of steps to reach the goal. These results confirm REVERB's reliability regarding force adjustment under a noisy environment. 
Fig.~\ref{fig_trajectory}(d)-(f) depict  the noise levels and number of selected $\SA$s over the whole observation space. When comparing Fig.~\ref{fig_trajectory}(d)-(e) to Fig.~\ref{fig_trajectory}(f), we notice that DT gathers more data as uncertainty increases and the agent approaches the target (position $\geq 0.45$). In scenarios where the noise level is high  but the agent is still far from the target (position $< -0.5$), requiring additional observations to improve control efficiency is considered ineffective. It is worth noting that this ineffectiveness does not have an adverse impact on the original task's performance.

Figure 4 illustrates different schemes' power consumption and Mean-root-mean-square-error (MRMSE). REVERB  proves to be the most efficient, consuming the least power and achieving the lowest MRMSE compared to other baselines. On average, REVERB consumes $1.09$ [W] to reach the goal, compared to $5.98$ [W]  of the \textit{Error-based} algorithm. Even though \textit{Traditional} one only queries 2 $\SA$s per QI, the power consumption is up to 5.4 [W] because this scheme takes more QIs to reach the goal, as in Fig.~\ref{fig_trajectory}(c). At the same time, REVERB's MRMSE is always maintained at approximately the same low level as the \textit{Cost-based} and \textit{Error-based} methods. 

The results achieved in Fig. 4 are explained by a snapshot of the uncertainty evolution and the strategic selection of $\SA$s in Fig. 5, based on their contributions to the DT's performance. It is noteworthy that the management of REVERB's uncertainty is subject to the control of both DT and RL requirements as flowing \eqref{glob_qos}. The requisites imposed upon the RL agent's behavior are primarily administered through the reward function delineated in \eqref{reward_shaping1}. Hence, DT only requests more observations when the DT threshold is surpassed or when the RL agent necessitates high precision, typically when the agent is nearing its goal and precise force control is imperative.

\begin{figure}[t]
\begin{minipage}{0.24\textwidth}
    %		\centering
    \vspace*{-10pt}
    \includegraphics[trim=0.cm 0cm 0cm 0cm, clip=true, width=1.8in]{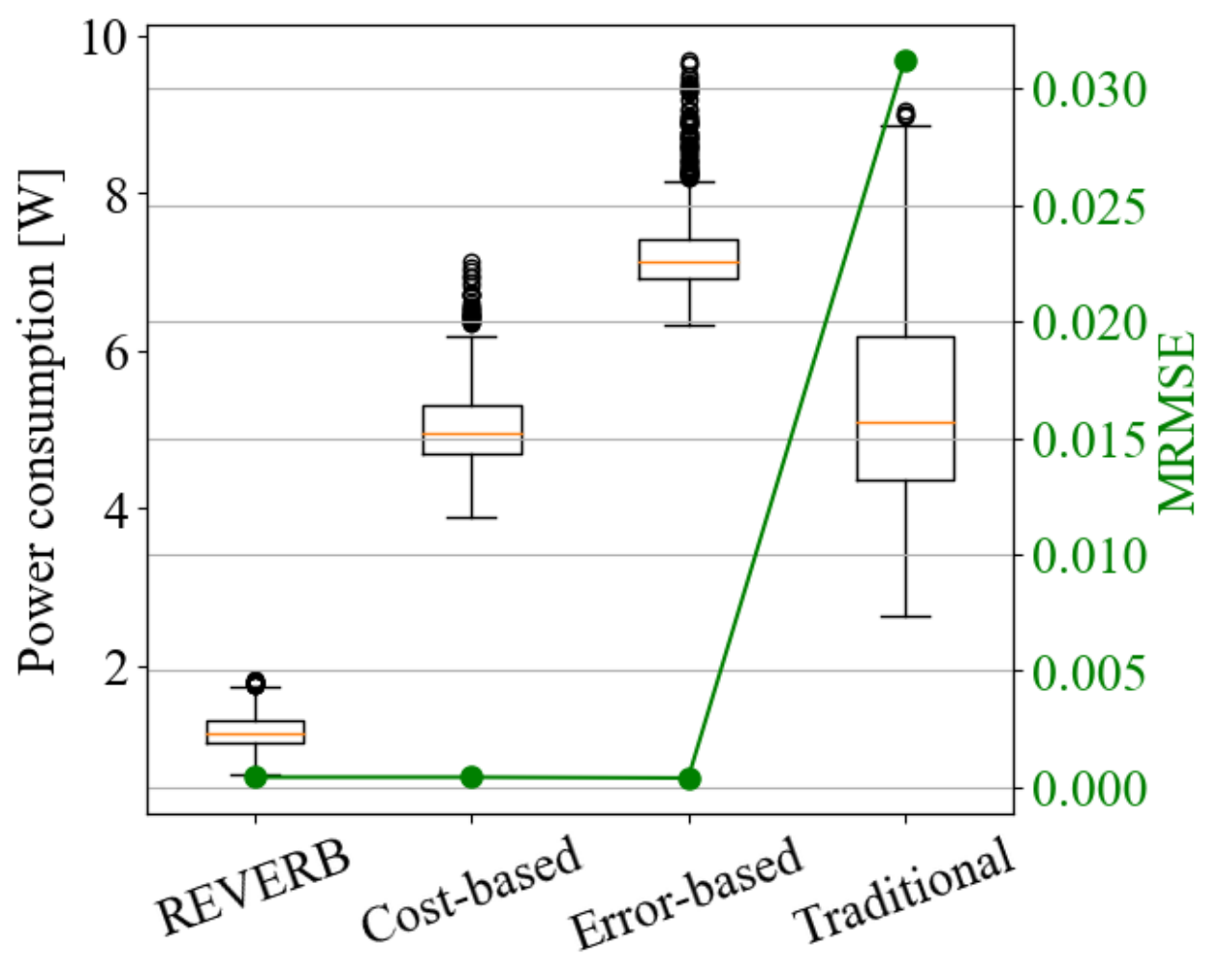} \\ 
    \centering {\footnotesize Fig. 4:  Power consumption \& MRMSE of different schemes.}
    \label{fig_powervsMRMSE}
    \vspace*{-10pt}
\end{minipage}
\begin{minipage}{0.24\textwidth}
    %		\centering
    \vspace*{-03pt}
    \includegraphics[trim=0.1cm 0cm 0.cm 0cm, clip=true, width=1.7in]{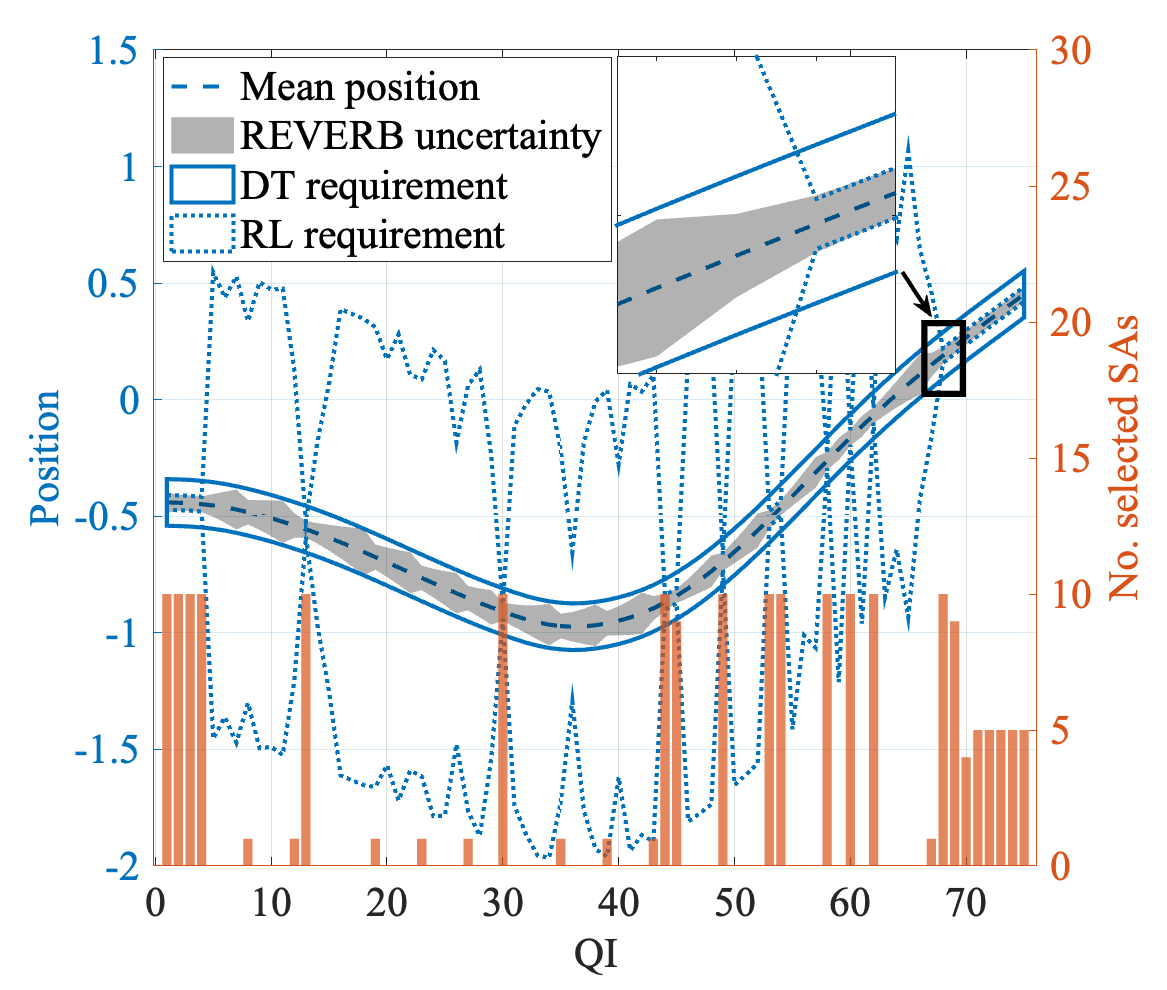} \\ 
    %		\vspace*{-15pt}
    \centering {\footnotesize Fig. 5: Snapshot of uncertainty evolution and no.  selected $\SA$s vs QI.} 
    \label{fig_uncertainty_evolution}
    \vspace*{-10pt}
\end{minipage}

%	\vspace*{-0.05cm}
%	\caption{Uncertainty level and number of scheduled $\SA$s.}
\label{}
\vspace*{-10pt}
\end{figure}

%%%%%%%%%%%%%%%%%%%%%%%%%%%%%%%%%%%%%%%%%%%%%%%%
\section{Conclusions}
%%%%%%%%%%%%%%%%%%%%%%%%%%%%%%%%%%%%%%%%%%%%%%%%
% \vspace*{-10pt}
This work introduced the DT framework for reliability monitoring, predicting and controlling of a communication system. Under the latency constraint, the  DT platform was shown to obtain more reliable control and trajectory tracking than conventional methods while significant saving communication cost. This result is achieved thanks to the proposed uncertainty control POMDP scheme combined with an efficient algorithm selecting subsets of partial $\SA$s to meet the requirements in the confidence of state estimation. Future study could explore the long-term impacts of scheduling decisions in complex systems and incorporate deep learning-based estimators.
\setstretch{0.9}
% \vspace*{-10pt}
\bibliographystyle{IEEEtran}
% \vspace*{-10pt}
% \balance
\bibliography{Journal}

\begin{thebibliography}{10}
\providecommand{\url}[1]{#1}
\csname url@samestyle\endcsname
\providecommand{\newblock}{\relax}
\providecommand{\bibinfo}[2]{#2}
\providecommand{\BIBentrySTDinterwordspacing}{\spaceskip=0pt\relax}
\providecommand{\BIBentryALTinterwordstretchfactor}{4}
\providecommand{\BIBentryALTinterwordspacing}{\spaceskip=\fontdimen2\font plus
\BIBentryALTinterwordstretchfactor\fontdimen3\font minus \fontdimen4\font\relax}
\providecommand{\BIBforeignlanguage}[2]{{%
\expandafter\ifx\csname l@#1\endcsname\relax
\typeout{** WARNING: IEEEtran.bst: No hyphenation pattern has been}%
\typeout{** loaded for the language `#1'. Using the pattern for}%
\typeout{** the default language instead.}%
\else
\language=\csname l@#1\endcsname
\fi
#2}}
\providecommand{\BIBdecl}{\relax}
\BIBdecl

\bibitem{tang2015tracking}
Y.~Tang \emph{et~al.}, ``Tracking control of networked multi-agent systems under new characterizations of impulses and its applications in robotic systems,'' \emph{IEEE Trans. Ind. Electron.}, vol.~63, no.~2, pp. 1299--1307, 2015.

\bibitem{9899718}
S.~Mihai \emph{et~al.}, ``Digital twins: A survey on enabling technologies, challenges, trends and future prospects,'' \emph{IEEE Commun. Surveys Tut.}, vol.~24, no.~4, pp. 2255--2291, 2022.

\bibitem{tse2005fundamentals}
D.~Tse and P.~Viswanath, \emph{Fundamentals of wireless communication}.\hskip 1em plus 0.5em minus 0.4em\relax Cambridge university press, 2005.

\bibitem{ruah2023bayesian}
C.~Ruah, O.~Simeone, and B.~Al-Hashimi, ``A bayesian framework for digital twin-based control, monitoring, and data collection in wireless systems,'' \emph{IEEE J. Sel. Areas Commun.}, 2023.

\bibitem{10092861}
V.-P. Bui, S.~R. Pandey, F.~Chiariotti, and P.~Popovski, ``Scheduling policy for value-of-information (voi) in trajectory estimation for digital twins,'' \emph{IEEE Commun. Lett.}, vol.~27, no.~6, pp. 1654--1658, 2023.

\bibitem{9768131}
F.~Chiariotti, A.~E. Kalør, J.~Holm, B.~Soret, and P.~Popovski, ``Scheduling of sensor transmissions based on value of information for summary statistics,'' \emph{IEEE Networking Lett.}, vol.~4, no.~2, pp. 92--96, 2022.

\bibitem{8017572}
M.~M. Azari \emph{et~al.}, ``Ultra reliable uav communication using altitude and cooperation diversity,'' \emph{IEEE Trans. Commun.}, vol.~66, no.~1, pp. 330--344, 2018.

\bibitem{NIPS1999_6449f44a}
V.~Konda and J.~Tsitsiklis, ``Actor-critic algorithms,'' \emph{Proc. NIPS}, vol.~12, 1999.

\bibitem{Kalman}
R.~E. Kalman, ``A new approach to linear filtering and prediction problems,'' \emph{J. Basic Eng.}, vol.~82, no.~1, pp. 35--45, 03 1960.

\bibitem{huang2019epkf}
Y.~Huang, W.~Yu, E.~Ding, and A.~Garcia-Ortiz, ``{EPKF}: Energy efficient communication schemes based on {Kalman} filter for {IoT},'' \emph{IEEE Internet Things J.}, vol.~6, no.~4, pp. 6201--6211, 2019.

\bibitem{simon2008digital}
M.~K. Simon and M.-S. Alouini, ``Digital communications over fading channels (mk simon and ms alouini; 2005)[book review],'' \emph{EEE Trans. Inf. Theory}, vol.~54, no.~7, pp. 3369--3370, 2008.

\bibitem{moore1990efficient}
A.~W. Moore, ``Efficient memory-based learning for robot control,'' University of Cambridge, Computer Laboratory, Tech. Rep., 1990.

\bibitem{sutton2018reinforcement}
R.~S. Sutton and A.~G. Barto, \emph{Reinforcement learning: An introduction}.\hskip 1em plus 0.5em minus 0.4em\relax MIT press, 2018.

\end{thebibliography}
%\vspace{-0.2cm}
\end{document}